# Coulomb blockade and hopping conduction in graphene quantum dots array

Daeha Joung[1,2], Lei Zhai[1,3], and Saiful I. Khondaker[1,2] *

[1] Nanoscience Technology Center, [2] Department of Physics, [3] Department of Chemistry,
University of Central Florida, Orlando, Florida 32826, USA.
* To whom correspondence should be addressed. E-mail: saiful@mail.ucf.edu

We show that the low temperature electron transport properties of chemically functionalized graphene can be explained as sequential tunneling of charges through a two dimensional array of graphene quantum dots (GQD). Below 15 K, a total suppression of current due to Coulomb blockade through GQD array was observed. Temperature dependent current-gate voltage characteristics show Coulomb oscillations with energy scales of 6.2-10 meV corresponding to GQD sizes of 5-8 nm while resistance data exhibit an Efros-Shklovskii variable range hopping arising from structural and size induced disorder.

## I.  INTRODUCTION

Reduced graphene oxide (RGO), a chemically functionalized atomically thin carbon sheet, provides a convenient pathway for producing large quantities of graphene via solution processing [1-5]. The easy processability of RGO and compatibility with various substrates including plastics makes them an attractive candidate for high yield manufacturing of graphene based electronic and optoelectronic devices. However, the electrical conductivity and field effect mobility values for RGO sheets are much inferior to that of pristine graphene [6-10]. This has been attributed to a large amount of disorder present in the RGO sheets. The structural characterization through Transmission Electron Microscopy (TEM), X-ray Photoelectron Spectroscopy (XPS), Scanning Tunneling Microscopy (STM) and Raman spectra show that RGO consists of ordered graphitic (nanocrystalline) regions surrounded by areas of oxidized carbon atoms, point defects, and topological defects [11-15]. The graphitic regions were estimated to be of 3-10 nm from the TEM, STM and Raman studies [11-15]. Optical studies of RGO also showed blue light emission [16] and infrared absorption [17, 18] determined by the size, shape and edge configuration $sp^2$ graphitic domain. All these studies clearly suggest that RGO should behave as a two dimensional array of graphene quantum dots (GQD), which should be verified from low temperature electron transport measurements. However, previous electrical transport studies of RGO in limited temperature range show 2D Mott variable range hopping (VRH) [11, 19-20] which is not expected from a QD array model. Additionally, Mott VRH neglects the Coulomb interaction between localized graphitic domains which may be significant at low temperature as recent study of individual 10 nm sized graphene quantum dots show room temperature Coulomb blockade (CB) [21, 22]. It is therefore quite puzzling why CB effect was not observed in low temperature transport of RGO sheets. A clear understanding of electron transport properties of RGO sheet is therefore still lacking which is of great significance for the developments of RGO as an important electronic and optoelectronic material.

In this paper, we present significant new understanding of the electron transport properties of RGO using low temperature electron transport measurements and show that the properties of RGO sheets can be described as a transport through an array of GQD with large size distribution (polydispersed array) where graphitic domains acts like QDs while oxidized domain behave like tunnel barriers between QDs. We show that below 15 K, the current is totally suppressed below a certain threshold voltage $V_t$ due to CB of charges through GQD array. For $V > V_t$, the current follow a scaling behavior $I \propto [(V-V_t)/V_t]^\alpha$ with $\alpha$ up to 3.4 expected from a quasi 2D QD array with topological inhomogenity. Current-gate voltage ($I$-$V_g$) curves at different temperatures show reproducible Coulomb oscillation corresponding to single electron tunneling which washes out between 70-120 K. These correspond to a charging energy of 6-10 meV giving a quantum dot size varying from 5-8 nm. Temperature dependent resistance data show ES type VRH ($T^{-1/2}$ behavior) arising from structural and size induced disorder with localization length of the same order with that of graphitic domain. Since the GQD size is tunable during the oxidation and reduction process, our study suggests that RGO will find many novel electronic and optoelectronic applications through tuning of GQD sizes.

## II. EXPERIMENTAL DETAILS

A. Synthesis of RGO

RGO sheets were synthesized through a reduction of GO prepared by modified Hummers method [23]. Oxidized graphite in water was ultrasonicated to achieve GO sheets followed by centrifugation for 30 minutes at 3000 rpm to remove any unexfoliated oxidized graphite. The pH of GO dispersion in water (0.1 mg/ml) was adjusted to 11 using a 5 % ammonia aqueous solution. Hydrazine hydrate was added and the solution was heated for 1 hour at 90 °C. The RGO suspension was spin coated on a mica substrate and examined using AFM. Figure 1(a) displays a tapping-mode AFM image of the RGO sheets along with their height (H) analysis. The lateral dimension (D) of our RGO sheets varies from 0.2 -1 μm. The line graph represents the thickness of the RGO sheets. Approximately 70 % of the sheets displayed a height of 1.0 ±0.2 nm.

B. device fabrication

Devices were fabricated on heavily doped silicon (Si) substrates capped with a thermally grown 250 nm thick $SiO_2$ layer. Source and drain electrode patterns of 500 nm x 500 nm (channel length x width) were defined by electron beam lithography (EBL) followed by thermal deposition of 5 nm thick Cr and 20 nm thick Au and standard lift-off. The RGO sheets were then assembled between the source and drain electrodes using AC dielectrophoresis (DEP) in a probe station [4]. DEP has been shown to assemble 2D, 1D and 0D nanomaterials at the selected position of the circuit for device applications [4, 24-27]. Figure 1 (b) shows a cartoon of the DEP assembly set up. A small drop of RGO solution was placed onto the electrodes pattern. An AC voltage of approximately 3 $V_{P-P}$ at 1 MHz was applied for 20-30 seconds after which the solution droplet was blown off by nitrogen gas. After DEP assembly, devices were thermally annealed in argon:hydrogen (1:3) gas at 200 $^0$C for 1 hour, details of which can be found in our previous study [4]. Figure 1(c) shows a tapping-mode AFM image of a representative device. From the thickness measurement, we estimate that 2 to 10 layers of RGO sheets have been assembled in the channel. In our previous publication, we reported that using DEP we can assemble RGO at selected position of the circuit with 100% device yield [4]. The devices were then bonded and loaded into a variable temperature cryostat for temperature dependent electronic transport measurements. Figure 1 (d) shows a schematic of the electrical measurement setup. The measurements were performed using a Keithley 2400 source-meter, and a current preamplifier (DL 1211) capable of measuring pA signal interfaced with LABVIEW program. A total of eight samples were investigated.

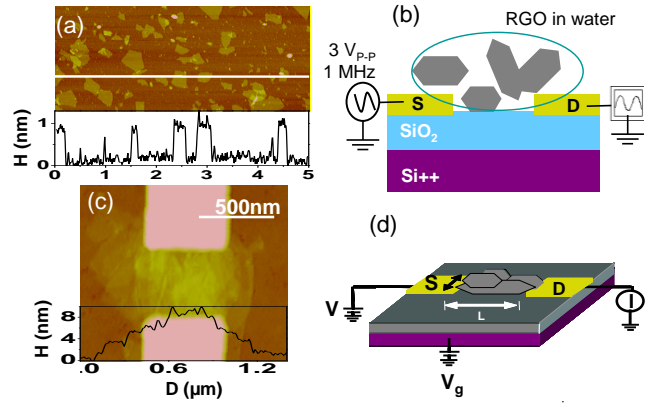

FIG. 1. (Color online) (a) Tapping-mode AFM images of RGO sheets with a height profile indicating majority of the sheets are single layer. (b) Cartoon of DEP assembly set-up. (c) Tapping-mode AFM of a RGO device assembled via DEP along with their height profile. The height (H) varies from 2 nm to 10 nm in the channel indicating that up to 10 layers of RGO sheets have been assembled in the channel. Scale bar represents 500 nm. (d) Cartoon of electronic transport measurement set-up.

### III. RESULTS AND DISCUSSION

Figure 2 (a) shows current-voltage (*I-V*) characteristics of a representative device at 30, 25, 20, 15, 10 and 4.2 K. The backgate voltage $V_g$ was kept fixed at 0 V. With decreasing temperatures, the *I-V* curves become increasingly nonlinear. However, all the curves are highly symmetric. As the temperature is lowered to less than 15 K, a complete suppression of current below a threshold voltage ($V_t$) was observed. Similar current suppression was observed in a previous study of individual GO devices with highly-asymmetric *I-V* curve and was explained by a Schottky barrier (SB) between metallic contact and GO [28]. However, our *I-V* curves are highly symmetric giving evidence that the current suppression and symmetric nonlinear behavior is not due to SB. Rather, such current suppression is due to CB of charges, as at low temperatures there is not enough energy for the charges to overcome Coulomb charging energies of the QD array formed by graphitic domains. In this scenario, the RGO sheet behaves as a GQD array where graphitic domains are quantum dots, and oxidized domains are tunnel barriers.

Theoretical studies of QD arrays by Middleton and Wingreen (MW) predicts that the *I-V* curves should follow the relation $I \propto [(V-V_t)/V_t]^\alpha$ for $V>V_t$, where $\alpha$ is the scaling exponent that depends on the dimensionality of the arrays [29]. Although, this theory was developed for nanocrystal arrays of uniform sizes (monodisperse array), however, experimentally it was found to be true for polydispersed array as well [30, 31]. Figure 2(b) shows $I$ plotted versus $(V-V_t)/V_t$ in a log-log scale using $V_t$ = 0.18, 0.24, 0.28 V at $T$ = 15, 10, and 4.2 K respectively. The symbols are the experimental data points while the solid lines are fits to the above equation. From the fits, we obtain $\alpha$ = 3.1, 3.3 and 3.4 at 4.2 K, 10 K, and 15 K respectively. For a two-dimensional array of nanoparticles, the theoretical value of $\alpha$ was predicted as 1.6 while numerical simulations yielded as 2.0 [29]. However, in previous experimental studies of two dimensional metal nanocrystal arrays, the exponent $\alpha$ was reported to vary from 2 to 2.5 which depends on size distribution, while for quasi 2D system with multilayered nanoparticles the value was 2.6 to 3.0 [30-34]. Although our system is a quasi-2D system, however, our $\alpha$ values are slightly higher than what has been reported. Recent experimental and computer simulations involving gold nanoparticles array with strong topological inhomogeneity show large scaling exponents $\alpha \approx 4.0$ [35, 36]. Since RGO has a lot of topological defects which came from oxidation and reduction process, the high value of $\alpha$ is in agreement with charge transport in an inhomogenous quasi 2D QD array network. In Figure 3, we show a schematic of RGO as a GQD array with strong topological inhomogeneity. The light gray areas represent GQDs, the white regions represent oxidized carbon groups and topological defects. This shows that the GQDs are isolated (or localized) by oxidized carbon atom and topological defect and there is a strong size distribution of GQDs. The lines between GQDs are indicates tunnel barriers.

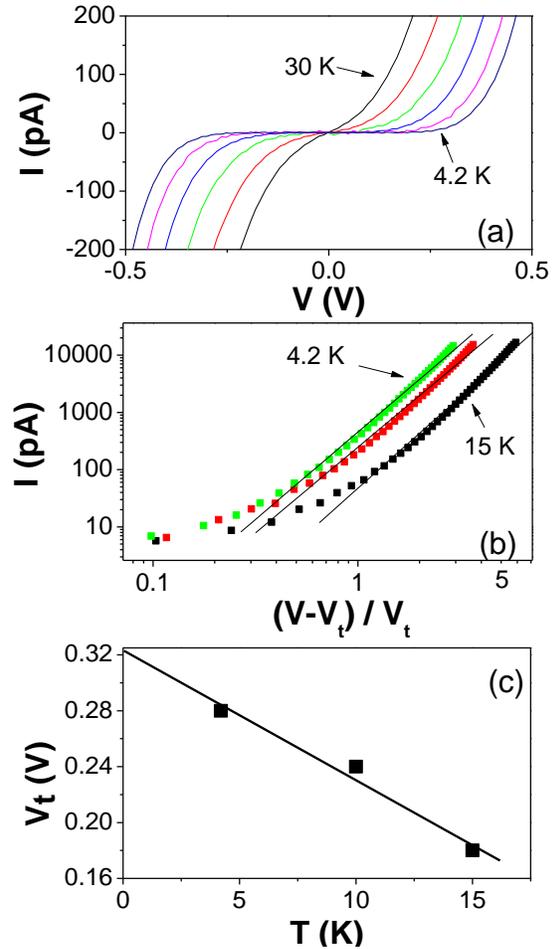

FIG. 2. (Color online) (a) Current(*I*)–voltage(*V*) characteristics of a representative RGO device at temperatures 30, 25, 20, 15, 10 and 4.2 K. Below 15 K, the current is zero for $V<V_t$ due to coulomb blockade of charges. Inset: AFM image of the device. Scale bar = 500 nm. (b) *I* vs. *(V-V_t)/V_t* curves plotted in a log-log scale. Slope of the curves gives the value of exponent $\alpha$ = 3.1, 3.3, and 3.4 at 4.2 K, 10 K, and 15 K respectively. (c) $V_t$ as a function of *T*. From the plot, $V_t$ (0) was estimated as 0.32 V.

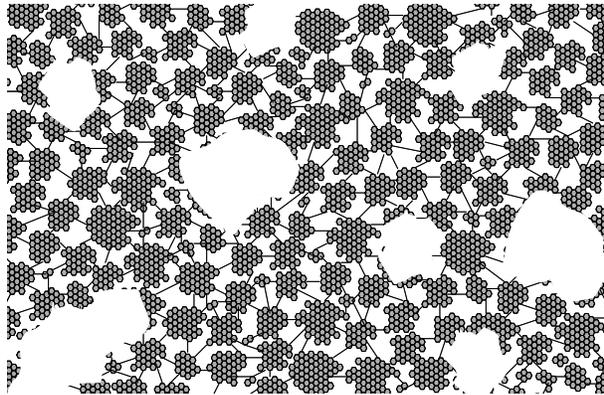

FIG. 3. Schemetic of RGO as GQD array. The light gray areas represent GQDs, the white regions represent oxidized carbon groups and topological defects. The lines between GQDs represent tunnel barriers.

Figure 2 (c) shows $V_t$ plotted versus $T$ from which we see that $V_t$ increases linearly with decreasing $T$. Extrapolation of the $V_t$ plot to 0 K provides the global threshold voltage $V_t(0)$ = 0.32 V . Similar $I$-$V$ curves were observed for all 8 samples with $\alpha$ varying from 2.52 to 2.80 and $V_t(0)$ varying from 0.32 to 0.42. For an array of nanoparticles of uniform size, $V_t(0)$ can be expressed as $V_t(0) \approx E_C(\beta N)$ where $E_c$ is charging energy of a QD, $\beta$ is a prefactor whose value depends on the dimensionality and arrays geometry (for a 2D array $\beta$ = 0.3), and $N$ is the number of QDs in the conduction path [29, 37]. From here, we can estimate the number of GQDs in our array contributing in the charge transport, however, we need to estimate $E_c$ first.

In order to calculate the $E_c$ of the GQDs, we measured $I$ as a function of gate voltage ($V_g$) at temperatures $T$ = 4.2 to 120 K. This is shown in Fig. 4. For clarity, the data in Fig. 4 a is plotted in a semi-log scale,with $I$ at 50, 60 and 70 K were divided by a factor of 1.5, 2, and 3.5 respectively. The reproducible peaks in $V_g$ correspond to single electron tunneling (Coulomb oscillations) through GQD arrays. The bias voltage was kept fixed at V = 0.3 V. The peaks in $V_g$ are not periodic, in agreement with the sequential tunneling of charges through multiple QDs. Such Coulomb oscillations have never been observed in previous studies of 2D metallic or magnetic QD array systems. This may be due to the fact that, the density of states (DOS) in those systems is higher and gate voltage has negligible effect in DOS. While in RGO, the DOS is low allowing the gate to tune the DOS giving rise to Coulomb oscillations. As the temperature is increased from 4.2 K to 120 K, two important features can be noticed. The peaks around $V_g = 0$, washes out around 70 K corresponding to a thermal energy of 6.2 meV. This is more clearly shown in Fig. 4 (b), where we plot $-10 < V_g < 10$ V regime up to $T = 70$ K of Fig. 4 (a). For clear presentation, curves from bottom to top were multiplied by a factor of 49, 41, 31, 23, 17, 12.5, 9, 6.7, 3.5, and 2.2 respectively. The other peaks survive up to 120 K (Fig. 4 c) which corresponds to a thermal energy of 10 meV.

From the semi classical orthodox theory of CB, the charging energy $E_C$ required to add an electron to a QD is given by, $E_c = e^2/2C_\Sigma$, where $C_\Sigma$ is the total capacitance which depends on the size of each QD and their inter-dot separation. In order to observe the coulomb oscillations, $E_c$ should be larger than thermal energy $k_B T$. Therefore our temperature dependent data gives an estimate of $E_c$ to vary from 6.2 – 10 meV. We suggest that this variation in charging energy is indicative of a large size distribution of GQDs in the transport pathway (polydispersed GQD array).

Using the $E_c$ values, the total capacitance is estimated to vary from $C_\Sigma = 8\text{-}13$ aF. Neglecting the size variation for the time being, $C_\Sigma$ can also be estimated from the geometrical consideration and can be written as $C_\Sigma = C_g + 9C$, where $C_g \approx 4\pi\varepsilon\varepsilon_o r$ and $C \approx 2\pi\varepsilon\varepsilon_o r \ln[(r+d)/d]$ are self-capacitance and mutual capacitance of QDs respectively [31] Here, $r$ is the radius of GQD, $2d$ is spacing between QDs, $\varepsilon$ is the dielectric constant of RGO, $\varepsilon_0$ is value for permittivity of vacuum, and the factor 9 is the average number of nearest neighbors of each QD in a quasi 2D system [31, 38]. The factor 9 was estimated as follows; For 2 D and 3 D hexagonal arrays, each nanocrystal has between 6 and 12 nearest neighbors. It was estimated in ref [31, 38] that for a quasi 2 D array, on average each GQD has ~ 9 nearest neighbors. The value of $\varepsilon_r$ can be calculated by using $\varepsilon_r = n^2 - k^2$, where $n$ and $k$ are refractive index and extinction coefficient of RGO film respectively. Using the values of $n$ and $k$ in the thermally reduced GO, the calculated value of dielectric constant $\varepsilon_r$ estimated to be around 3.5 [39]. By comparing experimental value and theoretical equation of $C_\Sigma$, we can calculate the value of $r = 2.5$ to 4 nm (domain size 5 – 8 nm) using $d = 0.75$ nm. These values obtained from electron transport spectroscopy is in excellent agreement with microscopic studies using TEM which highlighted that the size of graphitic regime varies from 3 to 10 nm [14-15]. We used $d = 0.75$ nm as the recent TEM study show that the typical size of oxidized or defective region varies from 1-2 nm [14] giving an average value for $2d = 1.5$ nm. The calculated domain size of GQD is also in good agreement with the domain size obtained from our Raman study [40].

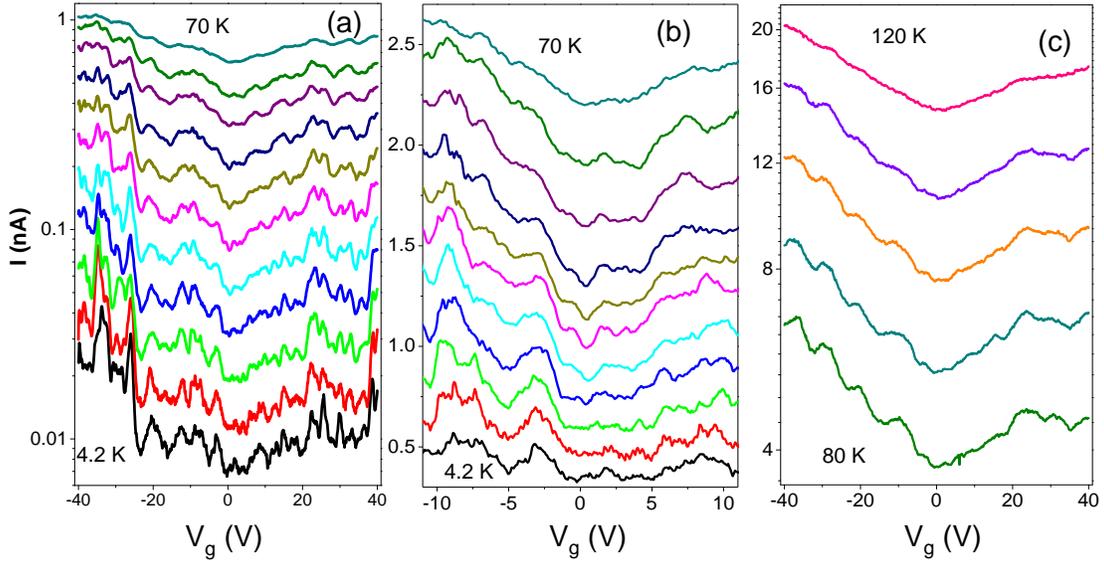

FIG. 4. (Color online) (a) Current (*I*) as a function of gate voltage ($V_g$) for $T = 4.2, 10, 15, 20, 25, 30, 35, 40, 50, 60$ and 70 K. The reproducible peaks correspond to Coulomb oscillations. For the unified view, *I* at 50, 60 and 70 K were divided by 1.5, 2, and 3.5. At 70 K, peaks around $V_g=0$ were washed out. This is more clearly shown in (b). For clarity, curves from bottom to top in (b) were multiplied by a factor of 49, 41, 31, 23, 17, 12.5, 9, 6.7, 3.5, and 2.2 respectively. (c) *I*-$V_g$ curves for $T = 80$ - 120 K with a step of 10 K. At 120K, all the oscillations were washed out. Bias voltage was 0.3 V.

We can now estimate the number of GQDs in the conduction pathway ($N$) of the array using the global threshold voltage formula $V_t(0) \approx E_C(\beta N)$. Using an average value of $E_c$ to be 8.1 meV, we estimate $N = 131$. This is a slight over estimation considering the average size of each dot as 6.5 nm and average inter-dot separation of 1.5 nm, we would expect about 65 QDs in 500 nm channel. The discrepancies may be due to the fact that we are using MW formula which neglects size disorder which can have a great influence in $V_t$. For example, if we consider that the smallest dot size has the most influence in the determination of $V_t$ as pointed out be Muller et al [41], then we obtain a more reasonable value of $N \sim 90$.

In order to further understand the electronic transport mechanism of the GQD array, we study the temperature dependence of the resistance of our devices. Temperature-dependence of resistance can provide evidence about size distribution and the degree of disorder of the GQD array. Figure 5 (a) shows the resistance ($R$) versus temperature ($T$) plot in the temperature range of 250-30 K for one of the devices. It can be seen that $R$ changes by over three orders of magnitude over this temperature range. $R$ was calculated by measuring the current at a constant $V = 100$ mV as the temperature was lowered (see solid line in Fig. 5a). We have also measured *I-V* curves at a few selected temperatures and obtained the $R$ values from the Ohmic part of the *I-V* curves (open circle in Fig. 5a). The $R$ values of the two measurements were in agreement. Below 30 K, the *I-V* curves were non-Ohmic

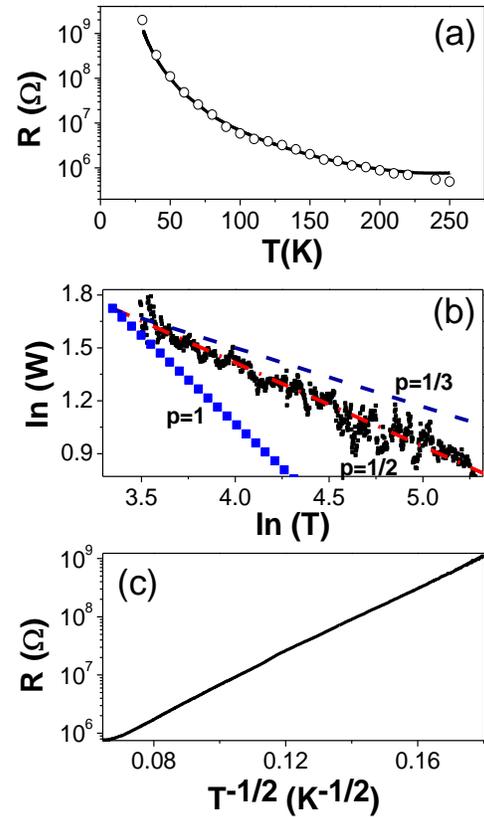

FIG. 5. (Color online) (a) Resistance ($R)$ versus temperature ($T$) in a semi-log scale showing four orders of change in $R$ for $T = 30$-$250$ K. The solid line respresntes R measured at a fixed V=100 mV as $T$ was decreased. The open symbols show $R$ measured from the Ohmic part of the *I-V* curves measured at a few selected $T$. (b) Reduced activation energy $W = -\partial lnR(T)/\partial lnT = p(T_0/T)^p$ plotted vs $T$ on a log-log scale. From the slope of this plot we obtain $p= 0.48 \pm 0.05$ corresponding to the ES-VRH. For a comparison we also show lines with $p = 1$ (activated hopping) and $p = 1/3$ (Mott VRH). Our data does not fit with those models. (c) $R$ in log scale as a function of $T^{1/2}$. From the slope we obtain $T_0 = 4200$ K.

under 100 mV and those data were discarded from this plot.

According to the QD array model, if QDs are monodispersed, the temperature dependence of resistance should follow thermally activated behavior $R \sim R_0 \exp(E_0/k_B T)$ [31], while if the nanocrystals have significant size variation (polydispersed) it should follow Efros-Shklovskii variable range hopping (ES-VRH), $R \sim R_0 \exp(T_0/T)^{1/2}$ [42, 43], where $T_0$ is a constant related to the disorderness of the material. Fitting resistance data with different behavior can be tricky and the same data can often fit with several behaviors (such as $T^1$, $T^{1/2}$, and $T^{1/3}$). A better way of determining the exponent value is to consider a generalized formula $R(T)=R_0 \exp(T_0/T)^p$ and then calculate the value of $p$ from $\ln W = A - p \ln T$, where $W = -\partial \ln R(T)/\partial \ln T = p(T_0/T)^p$ is the reduced activation energy and $A$ is a constant [44, 45].

Figure 5 (b) shows $\ln W$ plotted versus $\ln T$. From the slope (indicated by red line) of this curve, we obtain $p = 0.48 \pm 0.05$, which is consistent with ES VRH over the whole temperature range. For comparison, we have also plotted two lines for $p =1/3$ and $p =1$ which unequivocally show that the transport is only described by $p =1/2$ model. In previously reported data on single layer RGO devices, 2D Mott VRH ($p =1/3$) was reported [11, 19-20]. This may be due to limited temperature range of the data where it might be possible to fit the same data with both $T^{-1/2}$ and $T^{-1/3}$ law. Our self consistence analysis of the resistance data that span over three orders of magnitude clearly indicates that there is no conduction mechanism other than the $T^{-1/2}$ (ES VRH) for the whole temperature ranges. The characteristics of ES VRH model is in strong agreement with what is expected for a polydispersed GQD array.

The ES VRH indicates strong localization of wave functions in GQDs. Further analysis of ES VRH data allows us to calculate the localization length $\xi$ by plotting $R$ against $T^{-1/2}$ in a semi-log scale. This is shown in Fig. 5 (c), which shows a straight line as expected. From the slope of this curve, we obtain $T_0 = 4200$ K. $T_0$ is related to $\xi$ through $T_0 = [(2.8 e^2 / 4\pi\varepsilon\varepsilon_0 k_B \xi)$ [42]. The calculated value of $\xi$ is about 3.5 nm which is comparable to the calculated GQD sizes, indicating strong localization of the wave function inside each graphitic domain. Similar ES-VRH was observed for all the 8 samples with $\xi$ varying from 2.3-3.8 nm.

IV. CONCLUSION

All the measurements and analysis presented here clearly demonstrate that the low temperature charge transport properties of RGO can be modeled as due to a CB and hopping conduction through a polydispersed GQD arrays with topological inhomogenity. From our temperature dependence data, we obtain the GQD sizes to vary from 5 to 8 nm, in excellent agreement with previous TEM study. Observation of ES VRH with a localization length is comparable to the size of each GQD show that Coulomb interaction and size disorder play an important role. Our description of RGO sheet as a 2D GQD array suggests that RGO will find many novel electronic and optoelectronic applications through tuning of GQD sizes via controlled oxidation and reduction.

ACKNOWLEDGMENTS

We thank Eduardo Mucciolo for useful discussions. This work has been partially supported by U.S. NSF under Grant No. ECCS 0748091 to S.I.K. and under Grant No. DMR 0746499 to L.Z.